\documentclass[preprint2]{aastex}

\def\farcs{\hbox{$.\!\!^{\prime\prime}$}}
\def\fmag{\hbox{$.\!\!^{\rm m}$}}

\shorttitle{Decline of the space density of quasars}
\shortauthors{Vigotti et al.}

\begin{document}

\title{Decline of the space density of quasars between z = 2
  and z = 4}
\author{M. Vigotti}
\affil{Istituto di Radioastronomia del CNR, Via P. Gobetti 101,
I-40129 Bologna, Italy}
\email{vigotti@ira.cnr.it}
\author{R. Carballo\altaffilmark{1}}
\affil{Dpto. de Matematica Aplicada y C. Computacion,
  Universidad de Cantabria,
Avda. Los Castros s/n, E-39005 Santander, Spain}
\author{C.R. Benn}
\affil{Isaac Newton Group, Apartado 321, E-38700 Santa Cruz de La Palma, Spain}
\author{G. De Zotti}
\affil{INAF-Oss. Astr. Padova, Vicolo dell'Osservatorio 5, I-35122 Padova, Italy}
\author{R. Fanti\altaffilmark{2}}
\affil{Dipartimento di Fisica, Universit\`a  di Bologna, Via Irnerio 46, I-40126 Bologna, Italy}
\author{J.I. Gonzalez Serrano}
\affil{Instituto de Fisica de Cantabria (CSIC-UC),
Avda. Los Castros s/n, E-39005 Santander, Spain}
\author{K.-H. Mack\altaffilmark{3,4}}
\affil{Istituto di Radioastronomia del CNR, Via P. Gobetti 101,
I-40129 Bologna, Italy}
\author{J. Holt}
\affil{Department of Physics and Astronomy,
University of Sheffield, Hicks Building, Sheffield, S3 7RH, UK}


\altaffiltext{1}{Instituto de F\'{\i}sica de Cantabria (CSIC-UC), Avda. Los Castros s/n, E-39005 Santander, Spain}

\altaffiltext{2}{Istituto di Radioastronomia del CNR, Via Gobetti 101, I-40129 Bologna, Italy}

\altaffiltext{3}{ASTRON, Postbus 2, NL-7990 AA Dwingeloo, The Netherlands}

\altaffiltext{4}{Radioastronomisches Institut der Universit\"at Bonn, Auf dem H\"ugel 71, D-53121 Bonn, Germany}

\begin{abstract}
We define a new complete sample of 13 optically-luminous radio
quasars (M$_{AB}$(1450\AA)$< -26\fmag9$ and
$\log P_{1.4 GHz}$(W\,Hz$^{-1}$) $>$ 25.7) with
redshift $3.8 < z < 4.5$,
obtained by cross-correlating
the FIRST radio survey and the APM catalogue of POSS-I.
We measure the space density to be (1.0 $\pm$ 0.3) Gpc$^{-3}$,
a factor 1.9$\pm$0.7 smaller than the space density of
similar quasars at $z \approx 2$ (FBQS).
Using a new measurement of the radio-loud fraction of quasars
we find that at $z$ = 4 the total space density of quasars with
M$_{AB}$(1450\AA)$ < -26\fmag9$ is (7.4 $\pm 2.6) \rm ~Gpc^{-3}$.
This is a factor
1.8 $\pm$ 0.8 less than the space density at $z \approx 2$, found by the
2dF quasar survey.
This $z$ = 2 / $z$ = 4 ratio, consistent with that of the radio-loud quasars,
is significantly different from the ratio $\sim 10$ found for samples
including lower-luminosity quasars.
This suggests that the decline of the space density  beyond $z \approx 2$
is slower for optically-luminous quasars than for less-luminous ones.

\end{abstract}




\section{Introduction}

The space density of bright quasars is a factor
$\approx$ 100 higher at $z$ = 2 than at $z$ = 0.
Boyle et al. (2000; the 2dF survey) have shown that this increase can be
described in terms of pure luminosity evolution,
i.e. the shape of the luminosity function does not change with $z$,
up to at least $z = 2.3$.
At higher redshifts evidence for a decrease in space density,
consistent with pure luminosity evolution,
has been presented
by several authors (e.g. Kennefick et al., 1995 and references therein,
but see also Irwin et al. 1991).
The Sloan Digital Sky Survey (SDSS, Fan et al. 2001a) has pushed
exploration of the quasar phenomenon out to $z > 6$, the epoch of formation
of the first massive black holes (and galaxies) (Becker et al. 2001b).
The space density of  quasars at $z \approx 4$ with
M$_{AB}$(1450 \AA)
$ < -25\fm5$ ($\rm \Omega$ = 1 and $\rm \Lambda$ = 0,
H$_0$ = 50 km s$^{-1}$ Mpc$^{-1}$, as used throughout this paper)
was found by Fan et al. (2001b) to be more than one order of magnitude lower
than that found by Boyle et al. (2000) at $z\simeq 2.3$.
Fan et al. (2001b)
also found substantial deviations from pure luminosity evolution: the
bright end of the quasar luminosity function at $z>4$ appears to be
flatter than found at lower redshifts, i.e.
the decrease in space density with increasing redshift depends on
luminosity.

Radio-loud quasars are a small subset of the quasar population, but
as stressed by several authors (e.g. Peacock 1985; Dunlop \& Peacock
1990; Shaver et al. 1996, 1999), selection in the radio
drastically reduces
contamination by foreground stars with optical colors
similar to high-redshift quasars, and is unaffected by both intrinsic and
extrinsic dust obscuration and reddening.
Thus the high-redshift behaviour of
the radio-loud quasar luminosity function is a robust probe of
the abundance of massive objects at very early cosmological epochs. A second
important motivation for radio searches of high-$z$ quasars is
understanding the origin of the radio phenomenon. Differences, if
any, in the early evolutionary behaviour of radio and optical activity
would provide valuable clues.

Several searches for high-redshift radio-loud quasars
have been carried out
(e.g.: Shaver et al. 1996, 1999; Hook et al. 1998a;
Jarvis \& Rawlings 2000; Jarvis et al. 2001; Snellen et al. 2001;
Hook et al.,
2002).
Shaver et al. (1996, 1999) obtained complete redshift information
(i.e. no optical
magnitude limit) for a sample of 442 radio-loud flat-spectrum quasars with
flux densities $S_{2.7GHz} \geq 0.25\,$Jy.
They argued for a drop by more than a factor of 10 in space density from
$z\simeq 2.5$ to $z\simeq 6$, consistent with that found in the optical.

Hook et al. (1998b) reported indications that the peak in the
quasar space density may have occured at earlier cosmological epochs for
more radio-luminous sources, the decline in space density
with redshift being less pronounced for these radio sources.
Jarvis \& Rawlings (2000) analyzed the top decade in radio luminosity
(log $P_{2.7GHz}({\rm W\,Hz}^{-1}) > 28$, quasars and galaxies) of sources
from the Parkes Half-Jansky Flat-Spectrum Sample. They deduce a decline
by a factor $\sim 4$ between $z\simeq 2.5$ and  $z\simeq 5$, less abrupt
than envisaged by Shaver et al. (1996, 1999) and consistent with that
measured by Dunlop \& Peacock (1990) for radio sources in general.
Jarvis et al. (2001) investigated the high-redshift evolution of the
most-radio-luminous
($28.8 < \log P_{151MHz}({\rm W\,Hz}^{-1})<$ 29.8) steep-spectrum radio sources
using three
samples selected at low frequencies (3CRR, 6CE, and 6C$^\ast$). They
found the data consistent with a constant space
density between $z\simeq 2.5$ and  $z\simeq 4.5$ and exclude
a decline as steep as suggested by Shaver et al. (1996, 1999).
These results are not necessarily inconsistent, since
different ranges of radio luminosity and different frequency selections were
used.

No evidence for a rapid decline of the space
density of soft-X-ray  selected quasars at $z>2.7$ was apparent in the ROSAT
survey data analyzed by Miyaji et al. (2000).

In this paper we report the results of a search for $z \approx 4$
radio-loud quasars, down to radio flux densities  thirty  times fainter
than in previous searches,
and with no selection on radio spectral index.
The paper is organized as follows:

\noindent
In Sect. 2 we describe the construction of our sample
of $z$ = 4 quasars.

\noindent
In Sect. 3 we calculate the space density at $z \approx  4$, and
compare it with that at $z \approx 2$ derived from a sample of quasars of
similar optical and radio luminosity.

\noindent
In Sect. 4 we calculate the $z \approx 4$ space density of all quasars
optically brighter than M$_{AB}$(1450\AA) $ < -26\fmag9$, using
a new determination of the fraction of radio loud quasars at that redshift,
and compare it with that measured at $z$ = 2.

\noindent
Section 5 contains the discussion, and Sect. 6 our conclusions.

\section{Sample of $z \approx 4$ radio quasars}

\subsection{Candidate selection}
The broad-band optical color of quasars reddens dramatically
with redshift, as the Lyman-$\alpha$
forest and the 912-\AA ~Lyman-limit leave the blue band (Fig. 1).
To obtain a sample of high-redshift quasars, we therefore sought
optical identifications of FIRST radio sources
(Becker et al. 1995, White et al. 1997) with objects
catalogued by the Automated Plate Measuring facility (APM;
(McMahon \& Irwin 1992) as being star-like
on the POSS-I $E$ (red) plates,
with red magnitude $E \le18\fm8$, and with measured
color $ O - E \ge 3$, or $O$ fainter than the plate limit
($O$ $\sim 21\fm7$).

The FIRST radio catalogue (version of July 2000) includes
$\approx$ 660,000 radio sources  with peak flux density
S$_{1.4 GHz} > 1$ mJy in 2.24 sr and $7^h < RA < 17^h$,
$-5^\circ < DEC < 57 ^\circ$, of which
2.138 sr are covered by the APM catalogue.

We sought optical identifications lying closer than 1\farcs5
to each
radio position, corresponding to $\sim$ 3 standard deviations in the
expected combined radio and optical positional errors (see Fig. 2).

Searching to a larger radius would have netted 1\% more identifications,
but the contamination from foreground stars rises rapidly with
search radius (see again Fig. 2).
A similar search radius (1\farcs2) was used
by White et al (2000) in their search for First Bright Quasar Survey (FBQS)
candidates.
Our search yielded a starting list of 2323 candidates.

\begin{figure}
\epsscale{0.75}
\plotone{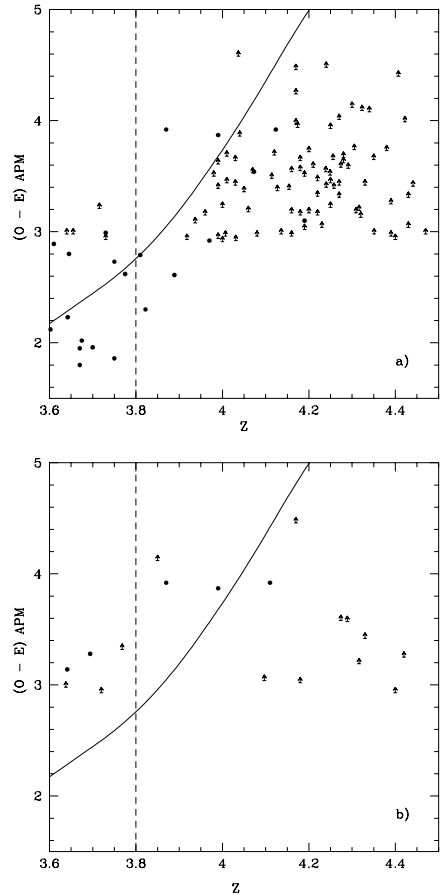}
\epsscale{1.0}
\vspace{-0.2cm}
\caption {POSS-I $O - E$ color vs redshift for high-redshift quasars
(a) with $3.6\leq z \leq 4.5$, $E\leq 18\fmag8$, from Djorgovski (2001, z $>$ 4)
    and from NED (as of May 2002, z $<$ 4);
(b) the 18 high-z FIRST quasars identified in our search
    (Section 2.2).
Filled symbols indicate measured $O - E$ colors, arrows are lower limits
(e.g. $>$ 3 for $E <$ 18\fmag7, $O$ limit 21\fmag7).
The quasars in (a) were selected in part on the basis of color, but
nevertheless trace the typical reddening of a quasar with
increasing $z$ as the continuum redward
of the Ly$\alpha$ line moves out of the observed $O$ band (red limit
$\approx$ 5000 \AA; prediction for typical quasar shown as solid curve,
computed
in a similar way as described in the Appendix).
As redshift increases through $z \approx$  4,
$O$-band shifts below the 912 \AA ~Lyman-limit and the
$O-E$ color rises rapidly.
As redshift increases above 4.5, the predicted brightness
of a quasar in $E$ band drops sharply, so few are likely to be detected.
The dashed line indicates $z$ = 3.8, above which we believe selection
by color to be nearly complete (Sect. 2.3).
\label{fig1}}
\end{figure}

Trial spectroscopy showed that $\approx$ 85\% of these `stellar' candidates
are faint galaxies.

In order to reduce contamination by these
galaxies, we ran the SExtractor program
(Bertin \& Arnouts 1996) on images of the 2323 candidates on the
POSS-II red plates, which have better sensitivity and spatial resolution
than POSS-I.
Eighty-two percent  of the candidates (1901 objects) were
clearly extended (i.e. galaxies).
Seventy-three candidates could not be found by SExtractor,
either because they were defects on the POSS-I plate (62 objects)
or due to confusion effects near a bright star (11 objects).

The remaining 349 candidates were sought in the
Minnesota APS catalogue of POSS-I (Pennington et al. 1993).
Fifty-five of the
candidates, undetected by APM on the $O$ plate (i.e.
potentially very red), were detected by APS.
These have $O-E <$ 3, so were removed from the sample.
This left 294 candidates, 194 of which we classified,
on the basis of the SExtractor-generated parameters,
as `star-like', and 100 as `possibly star-like'.
CCD images of a third half of these were obtained in 1\farcs3-seeing
with the Loiano 1.5-m telescope of the Astronomical Observatory of Bologna.
On these images, 25 of the 70 `star-like',
and 28 of the 33 `possibly star-like' candidates were extended.
We have not investigated further the 100 `possibly star-like' objects.
As a result we expect an incompleteness of $\approx  100 \times 5/33
= 15 (\pm 4)$ stellar candidates.

\subsection{Spectroscopy}
Spectra were obtained of 121 of the 194 candidates,
using the IDS spectrograph of the Isaac Newton Telescope
(Benn et al. 2001, Holt et al., in prep.).
Eighteen of the 121 (15\%) are quasars with  $3.6 < z < 4.5$.
The fraction of
high-redshift quasars varies little with
the optical magnitude of the candidates,
but depends strongly on radio flux density:
$(7.7 \pm 2.9) \%$ for $S_{1.4GHz} <$ 10 mJy and $(37 \pm 11)\%$ for
$S_{1.4GHz} > 10$ mJy.
Amongst the 73 candidates not observed spectroscopically, there are 6 with
$S_{1.4GHz} > $ 10 mJy and 67 with $S_{1.4GHz} < 10$ mJy.
On the basis of the above high-redshift-quasar fractions,
and accounting also for the $\sim$ 15 missed
stellar candidates (Sect. 2.1), we expect to have missed
a total of
$9 \pm 2$ high redshift quasars, i.e. the search is
$\approx (67 \pm 5) \%$  complete or, equivalently,
corresponds to an area on the sky
of 1.432 sr.

\begin{figure}
\epsscale{1.0}
\plotone{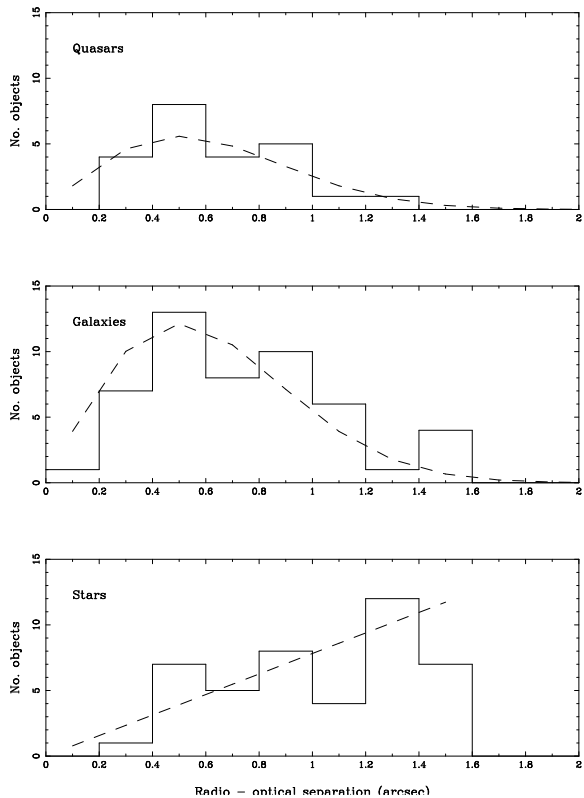}
\epsscale{1.0}
\caption {The distributions of radio-optical separations $r$ for
the star-like candidates classified spectroscopically
as quasars, galaxies and stars (total 194 objects).
For quasars and galaxies, the dashed curves are of the form $r\cdot exp(-r^2/2\sigma^2)$, with
$\sigma$ = 0\farcs5.
For stars, the dashed straight line indicates the form expected for
misidentifications with foreground stars.
\label{fig2}}
\end{figure}

\subsection{Sample completeness}
Our search for high-redshift quasars within this area will be incomplete
for a number of other reasons, that we discuss below:

\medskip

1) {\it Incompleteness of the APM catalogue}. According to McMahon et al.
(2001), APM classifies correctly 92\% of $E < 19^m$  objects on at
least one of the $O$ or $E$ plates. Assuming that the APM $E$ and $O$ plate
misclassifications are independent and occur at a similar rate,
then a 92\%-correct classification on at least one plate implies 72\%-correct
classification on individual plates. This last figure is a lower limit on
the incompleteness of our $E$ plate based star-like classification
(the classification on the $O$ and $E$ plate cannot be completely
independent).

For an independent estimate of our incompleteness, we searched
for APM detections of 100 $z >$ 4 quasars with $E \le 19^m$ from the list
of Djorgovski (2001), which were
originally found either on the POSS-II plates or in
SDSS (CCD based).
All 100 were detected by APM.
Eighteen of these quasars appeared extended in
the APM $E$ band, i.e. the search was $(82 \pm 4)\%$ complete.

As a further check we searched the APM catalogue for counterparts of 4949
objects with $ 16^m \le E \le 19^m $ catalogued as star-like
in the APS.  APM classifies as extended $(13 \pm 0.5) \%$ of these
objects and misses a further $(3 \pm 0.3)\%$, i.e. is 84\% complete.
Because the APS classification algorithm (neural
network) is believed to be more reliable than that of APM,
we assume that disagreements about classification by APM and APS are
due mainly to errors in the former.

To check the APS classification reliability, we constructed a sample of 190
quasars,
140 from the FBQS survey with $ 16^m < E < 18^m$ and 50 from the B3VLA QSS
survey (Vigotti et al. 1997) with $18^m < E < 19^m$.
Only 10 quasars
were classified extended by APS, i.e. an incompleteness of 5\%.
This is three times smaller than the APM incompleteness, consistent
with our assumption above.

The above two estimates of APM completeness are in
good agreement and we assume a value of
$(84 \pm 2$) \%.
In addition, a few star-like objects might have been misclassified
extended when we ana\-lysed the images with SExtractor, or at the
stage of CCD imaging, but the number is likely to be very small.

\smallskip

2) {\it Omission of quasars with extended radio emission}.
Optical counterparts were sought at the positions of
individual FIRST radio
sources, but not at the mid-points of possible double sources.
In the First Bright Quasar Survey (FBQS)
 in the South Galactic Cap, Becker et al. (2001a)
also sought counterparts along the lines
joining possible double sources of separation up to 30$''$,
and found that a search at the positions of
individual radio components is 96\% complete.
At $z \sim$ 4, the rest-frame frequency of observation
is even higher than
for FBQS, favouring more flat-spectrum compact sources.
We therefore expect our sample to be more complete,
and we assume a completeness of (98 $\pm$ 1)\%.

\smallskip

3) {\it Incompleteness due to errors in the radio-optical positions}.
The distribution of the radio - optical separations of the quasars
(Fig. 2) is consistent with a rms error of 0\farcs5,
implying that our search out to radius 1\farcs5 is
$\approx 99\%$ complete.

\smallskip
4) {\it Incompleteness due to color selection}.
The fraction of $z >$ 3.8 quasars with $O-E >$ 3$^m$ can be estimated
approximately from the color distribution in Fig. 1a (taken from
the literature).
81 of the 85 quasars with $z >$ 3.8
have $O-E >$ 3, i.e. this color criterion
finds ($95 \pm 1.5)$ \%
of quasars with  z$>$ 3.8, but the efficiency
falls rapidly with decreasing redshift.
We therefore restrict our sample to $z > 3.8$.

\medskip

Combining the above, the completeness of our search for
$z >$ 3.8 quasars in FIRST is
$\approx (77 \pm 6)$\% over the effective area (Sect. 2.2)
of 1.432 sr.

\smallskip

5) {\it Incompleteness of the FIRST radio survey}.
   The FIRST catalog is severely incomplete at faint flux densities.
   From Prandoni et al. (2001) we estimate a completeness of $\approx
   50 \%$ for $S \le 1.25$ mJy and of $\approx 75 \%$ for $1.25~{\rm mJy}
   \le {\rm S} \le 1.5$ mJy. This flux-density-dependent incompleteness
   is in addition to that discussed above.
    
\smallskip

\subsection{The $z \approx 4$ radio-loud sample}
The 13 radio-loud quasars with $z \geq 3.8$, the $Q_{z4RL}$ sample, are
listed in Tab. 1.

The  $E$ magnitudes were recalibrated using APS catalogue (rms error 0\fmag2,
better than APM, because the APS photometry was
calibrated on a plate-by-plate basis) and corrected for Galactic extinction.
After re-calibration the limiting magnitude of the sample is $E = 19^m$

The epoch-1950 absolute magnitudes in the rest-frame UV continuum,
$M_{AB}$(1450 \AA), were obtained from the $E$ magnitudes,
using a template quasar spectrum
(the procedure is described in the Appendix).
The 13 quasars are optically very luminous,
M$_{AB}$(1450 \AA)$ \le -26\fmag9$.

The radio luminosities were calculated using a median spectral index
$\alpha_R = - 0.3 $, rms = 0.4  (S$_{\nu} \propto \nu^{\alpha}$),
and range $ \rm \log P_{1.4GHz}(W\,Hz^{-1}) \ge$ 25.7.
The median spectral index was obtained via 5-GHz observations
of some sources using
the Effelsberg radio telescope (Holt et al. 2003, in prep.).

\begin{small}
\begin{table*}[t]
\centering
\caption{Properties of the $z$ = 4 radio-loud sample \label{Tab-1}}
\begin{tabular}{cccrcrcccc}
\tableline\tableline
 RA & Dec& $E$ &  $O-E$ & $z$ & S$_{1.4 GHz}$ & logP$_{1.4 GHz}$ & M$_{AB}$ & $z_{max}$ & $\rho$\\
\multicolumn{2}{c}{J2000}&&&&mJy&WHz$^{-1}$&$(1450\AA)$&& $Gpc^{-3}$ \\
07 25 18.29& +37 05 17.9 &  18.7& $>$ 3.4&  4.330&  26.6&  27.11& $-26.9$&  4.379&  0.0517\\
07 47 11.16& +27 39 03.6 &  17.8& $>$ 4.5&  4.170&   1.1&  25.70& $-27.6$&  4.366&  0.0530\\
08 31 41.68& +52 45 17.1 &  15.8&     3.9&  3.870&   1.3&  25.72& $-29.8$&  4.500&  0.0432\\
08 39 46.11& +51 12 03.0 &  18.9& $>$ 3.2&  4.420&  41.6&  27.32& $-27.3$&  4.438&  0.0472\\
09 18 24.38& +06 36 53.5 &  19.0& $>$ 3.0&  4.180&  26.5&  27.09& $-26.9$&  4.206&  0.0731\\
09 41 19.36& +51 19 32.3 &  18.0& $>$ 4.1&  3.850&   2.5&  26.00& $-27.6$&  4.470&  0.0450\\
10 53 20.42&$-00$ 16 50.4 & 17.8& $>$ 3.6&  4.289&  13.8&  26.82& $-27.7$&  4.483&  0.0441\\
10 57 56.27& +45 55 53.0 &  16.9&     3.9&  4.110&   1.1&  25.70& $-28.5$&  4.350&  0.0544\\
12 11 34.41& +32 26 15.8 &  18.2& $>$ 3.0&  4.097&   3.7&  26.21& $-27.2$&  4.421&  0.0484\\
13 09 40.61& +57 33 09.1 &  18.3& $>$ 3.6&  4.274&  11.3&  26.73& $-27.1$&  4.406&  0.0495\\
13 25 12.52& +11 23 29.9 &  18.6& $>$ 2.9&  4.400&  71.1&  27.55& $-27.5$&  4.455&  0.0460\\
14 23 08.19& +22 41 57.5 &  18.7& $>$ 3.2&  4.316&  35.4&  27.23& $-26.9$&  4.377&  0.0519\\
16 39 50.51& +43 40 03.6 &  17.7&     3.9&  3.990&  25.2&  27.03& $-27.8$&  4.500&  0.0432\\
\end{tabular}
\addtocounter{table}{-1}
\caption{The columns give:
(1 - 2) J2000 optical coordinates,
(3) $E$ magnitudes recalibrated using APS;
(4) APM $O - E$ color;
(5) spectroscopic redshift;
(6) FIRST radio flux density (peak or integrated, see next column),
(7) radio luminosity, calculated using the peak flux density for S$_{1.4GHz} < 10$ mJy
and the integrated flux density for S$_{1.4GHz} > 10$ mJy, as recommended
by Ivezic et al. (2002, see their Fig. 4), and
K-corrected using the median radio spectral index
$\alpha_R=-0.3$
(S$_{\nu} \sim \nu^{\alpha}$, as used throughout this paper),
(8) absolute magnitude $M_{AB}$ at rest-frame 1450 \AA\ (see Appendix),
(9) maximum redshift out to which the quasar could have been detected (i.e.
    with  $E <$ 19\fmag0, $S_{1.4GHz} >$ 1 mJy),
(10) contribution of this source to space density, assuming
effective area 1.432 sr, but without correction for the
incompletenesses discussed in Sect. 2.3 (see text).
}
\end{table*}
\end{small}
\begin{figure}
\epsscale{.9}
\plotone{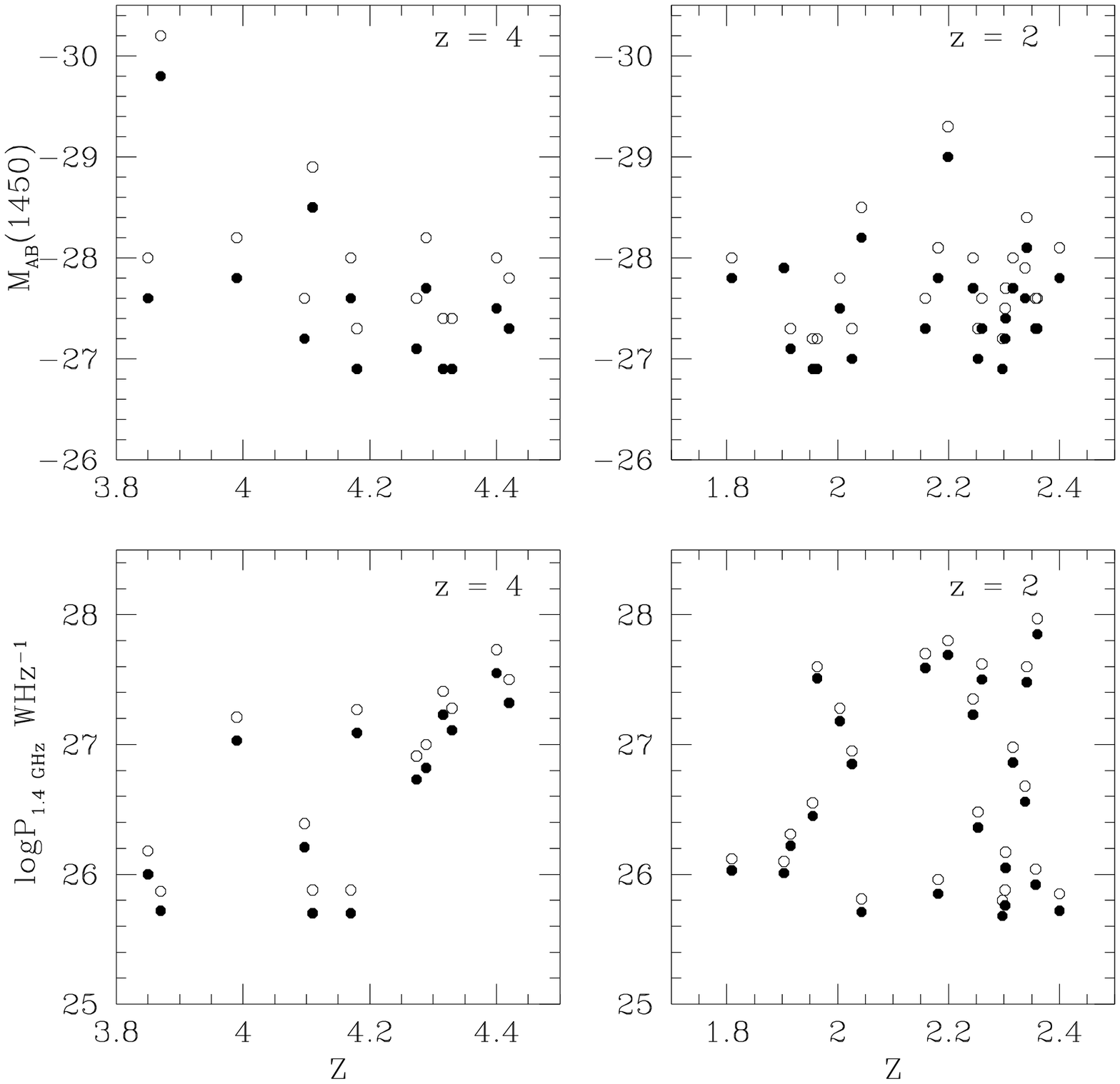}
\epsscale{1.0}
\caption {The plots of Radio and Optical luminosity vs z,
for the $Q_{z4RL}$ sample (left plots) and
for the $Q_{z2RL}$ sample (right plots).
Filled symbols are for the Einstein -- de Sitter cosmology, empty symbols
are for the $concordance ~model$ (see Sect. 3.2).
\label{fig3}}
\end{figure}

As a check of our method of deriving $M_{AB}$(1450 \AA),
we made, for 8 of the quasars, independent measurements based
on recent $I$-band photometry
obtained during service time at the 1.0-m
Jacobus Kapteyn Telescope in La Palma (Holt et al., in prep.).
At $z$ = 4.2, $I$ band corresponds to rest-frame wavelength
$\sim$ 1540\AA,
close enough to the reference wavelength 1450\AA  ~that we can
K-correct using the median optical spectral index
$\alpha_{OPT}$ = $-0.7$ from Fan et al. (2001c).
For two of the 8 quasars, 0831+52 and 1423+22,
we found a large difference between
the $M_{AB}$ values derived using these two methods, the $I$-photometry
giving $M_{AB}$
respectively dimmer by 0\fmag4 and brighter by 1\fmag0.
These differences are probably
due to intrinsic changes in luminosity between epochs 1950 and 2002.
For the remaining 6 quasars the mean difference is  0.0 $\pm$ 0.1,
suggesting that any systematic error in our M$_{AB}$ magnitude scale is
$\la$ 0\fmag1.

The mean radio and optical luminosities of the sample are: $\langle log P_{1.4GHz}\rangle =
26.6 \pm 0.2$ and $\langle M_{AB}\rangle = -27\fmag6 \pm 0\fmag2$,  the median values being
26.8 and $-27\fmag5$ respectively.

\section{Decline of the space density of radio loud quasars
between $z = 2$ and $z = 4$}

\subsection{Comparison sample of $z \approx 2$ radio quasars}

For a sample of radio quasars with comparable radio and optical luminosities,
but at $z \approx$ 2, we draw on the 0.817-sr First
Bright Quasar Survey (FBQS, White et al. 2000).
FBQS uses $O$,~$E$  magnitudes recalibrated using APS, with a limiting magnitude
$E = 17\fm8$.
Of the original 1238 quasar candidates, 636 were confirmed quasars.
No spectrum is available for 108 of the 1238 candidates, i.e. the
spectroscopy is 91\% complete, corresponding to an effective area
on the sky of 0.743 sr.

Of the 45 FBQS quasars with $1.8 \le z \le 2.4$, 23 have
luminosities in the range of the $Q_{z4RL}$ sample, i.e.
$\log P_{1.4~GHz} \ge 25.7$ and M$_{AB}$(1450 \AA)$ \le -26\fmag9$.

The mean radio and optical
luminosities values are $\langle \log P_{1.4GHz}\rangle = 26.6 \pm 0.2$ and
$\langle M_{AB}\rangle = -27\fmag5 \pm 0.1$.
The median values are 26.4 and $-27\fmag4$
respectively (consistent, within the statistical noise, with those of the z = 4
sample).
We thus have a lower-redshift sample directly comparable with
our high-redshift one.
Fig. 3 shows the distributions of the sources in
radio and optical luminosity and redshift
for each
sample, for the Einstein -- de Sitter cosmology and for the $concordance
~model$ (see Sect. 3.2)

This sample (Tab. 2) will be referred as the $Q_{z2RL}$ sample.
The radio luminosities $P_{1.4GHz}$ were calculated assuming a
radio spectral index  $\alpha_R = -0.3$.  This is the median spectral
index obtained for a subsample with
GB6 4.8-GHz flux densities (Gregory  et al. 1996), and is similar to that of
the $z \approx$ 4 sample.
The absolute magnitudes $M_{AB}$(1450 \AA) have been calculated from the
$E$ magnitudes using a K-correction = 2.5 log$(1+z)^{(1+\alpha_{OPT}) }$ and
$\alpha_{OPT} = -0.5$, that of the FBQS composite spectrum
(Brotherton et al. 2001).

As an independent check, we also calculated  the absolute magnitudes
using the quasar template (following a procedure similar to that
used at $z$ = 4, see Appendix), obtaining
similar results, within $\pm 0\fmag1$.

This sample will suffer from some of the same incompletenesses as
the $Q_{z4RL}$ sample (Section 2.3).
The search for star-like objects $E\leq$ 17\fmag8 in the APM
will be $\approx$ ($96 \pm 1$)\% complete.
An additional 4\% of objects will be missed due to not searching at the
mid-points of possible double radio sources (Becker et al. 2001a).
No losses are expected due to
color selection, nor for incompleteness of the FIRST at faint flux densities
(these sources are brighter than 3 mJy).
In total, the $Q_{z2RL}$ sample of quasars
is expected to be ($92 \pm 1.4$)\% complete over 0.743 sr.

\begin{small}
\begin{table*}[t]
\centering
\caption{Properties of the $z$ = 2 radio-loud sample \label{Tab-2}}
\begin{tabular}{cccccrcccc}
\tableline\tableline
 RA & Dec& $E$ &  $O-E$ & $z$ & S$_{1.4 GHz}$ & logP$_{1.4 GHz}$ & M$_{AB}$ & $z_{max}$ & $\rho$\\
\multicolumn{2}{c}{J2000}&&&&mJy&WHz$^{-1}$&$(1450\AA)$&& $Gpc^{-3}$ \\
07 29 28.47& +25 24 51.99&  17.2&   0.6&   2.303&    6.8&  26.05&  $-27.4$&  2.400&  0.0761\\
07 58 13.94& +26 24 18.75&  17.1&   0.7&   1.915&   13.9&  26.22&  $-27.1$&  2.400&  0.0761\\
08 04 13.67& +25 16 34.07&  17.4&   0.7&   2.302&    3.5&  25.76&  $-27.2$&  2.400&  0.0761\\
08 57 26.95& +33 13 17.24&  17.1&   0.3&   2.338&   21.3&  26.56&  $-27.6$&  2.400&  0.0761\\
09 04 44.33& +23 33 54.05&  16.9&   0.3&   2.244&  107.3&  27.23&  $-27.7$&  2.400&  0.0761\\
09 10 54.18& +37 59 15.11&  17.2&   0.0&   2.158&  265.9&  27.59&  $-27.3$&  2.400&  0.0761\\
10 12 11.43& +33 09 25.96&  17.3&   0.0&   2.260&  196.3&  27.50&  $-27.3$&  2.400&  0.0761\\
10 35 10.98& +35 10 19.51&  17.3&  -0.2&   1.955&   22.9&  26.45&  $-26.9$&  2.400&  0.0761\\
10 36 41.94& +25 02 36.63&  16.9&   0.7&   2.004&  118.5&  27.18&  $-27.5$&  2.400&  0.0761\\
10 54 27.15& +25 36 00.33&  16.9&   1.4&   2.400&    3.0&  25.72&  $-27.8$&  2.400&  0.0761\\
11 22 41.46& +30 35 34.88&  16.4&   0.4&   1.809&   10.0&  26.03&  $-27.8$&  2.400&  0.0761\\
12 25 27.38& +22 35 12.72&  16.2&   0.2&   2.043&    3.9&  25.71&  $-28.2$&  2.400&  0.0761\\
12 28 24.98& +31 28 37.74&  15.5&   0.4&   2.199&  323.8&  27.69&  $-29.0$&  2.400&  0.0761\\
13 24 22.53& +24 52 22.25&  17.4&   1.5&   2.357&    4.9&  25.92&  $-27.3$&  2.400&  0.0761\\
13 45 20.42& +32 41 12.59&  17.6&   0.3&   2.253&   14.6&  26.36&  $-27.0$&  2.400&  0.0761\\
14 15 28.47& +37 06 21.49&  17.4&   0.9&   2.360&  409.7&  27.85&  $-27.3$&  2.400&  0.0761\\
14 16 17.38& +26 49 06.20&  17.7&   0.8&   2.297&    2.9&  25.68&  $-26.9$&  2.400&  0.0761\\
14 21 08.72& +22 41 17.42&  16.8&   0.1&   2.181&    4.7&  25.85&  $-27.8$&  2.400&  0.0761\\
14 23 26.07& +32 52 20.34&  16.3&   0.2&   1.903&    8.7&  26.01&  $-27.9$&  2.400&  0.0761\\
14 32 43.32& +41 03 28.04&  17.4&   0.7&   1.963&  261.7&  27.51&  $-26.9$&  2.400&  0.0761\\
16 03 54.15& +30 02 08.88&  17.4&   0.7&   2.026&   54.2&  26.85&  $-27.0$&  2.400&  0.0761\\
16 34 12.77& +32 03 35.45&  16.6&   0.3&   2.341&  176.5&  27.48&  $-28.1$&  2.400&  0.0761\\
16 51 37.56& +40 02 18.71&  16.9&   0.3&   2.316&   43.9&  26.86&  $-27.7$&  2.400&  0.0761\\

\end{tabular}
\addtocounter{table}{-1}
\caption{The columns are as for Tab. 1, except that in column:
(4) $O - E$ is the APM color using APS recalibration;
(8) M$_{AB}$(1450\AA) is calculated from the $E$ magnitude
    (the $O$ magnitude is closer in rest-frame wavelength, but may
    include Ly$\alpha$ in the wing of the filter response);
(9) the maximum redshift out to which the quasar could have been detected
    is calculated assuming
    $E <$ 17\fmag8, $S_{1.4GHz} >$ 1 mJy;
(10) effective area 0.743 sr is assumed.
}
\end{table*}
\end{small}

\subsection{The space density of radio-loud quasars at $z \approx 4$ and
at $z \approx 2$ }

We have two shells of volume: one defined by  $1.8\le z\le2.4$
(which has been chosen to yield a good statistical sample around $z = 2$)
and the other by  $3.8 \le z \le 4.5$ (determined by the fact that
at $z$ larger than 4.5 the predicted brightness of a quasar
in $E$ band drops sharply, so few are likely to be detected).
For each of the two samples of quasars,
the space density was calculated as follows.
For each source we computed $z_{max,r}$  and $z_{max,o}$ the maximum redshifts
at which the source would reach the radio flux-density limit
or the optical magnitude limit of the sample.

In addition, for the $z = 4$ sample of quasars, we calculated the
{\it minimum} redshift at which the quasar could have been seen, since
the K-correction could in principle cause the quasar to dim with decreasing
redshift (Fig. 6).
For all the sources we found this minimum redshift to be $<3.8$.

The maximum redshift $z_{max}$ at which the quasar could have been
included in the sample is the minimum of $z_{max,r}$ and $z_{max,o}$.

Just two of the quasars in the $z = 4$ sample
are $radio~ limited$ ($z_{max,r} \le z_{max,o}$),
while in the $z =$ 2 sample there are none.
The uncertainty in $z_{max,r}$ due to the assumption
of an average radio spectral index ($-0.3$) is $\sim 0.1$
and implies a negligible effect on the calculated space density
($ < 2\%$).

The lower limit on redshift is the sample limit $z_0$ (1.8 for the
$z \approx$ 2 sample, 3.8 for the z $\approx$ 4 sample).
The space density contributed by each quasar is the
inverse of the volume of a shell
bounded by  $z_0$ and $z_{max}$, and with surface area equal to the
effective area covered by the sample (0.743 and 1.432 sr
for the $Q_{z2RL}$ and the  $Q_{z4RL}$ samples respectively).
This volume is:

\noindent
$ V = k(( 1 - \frac{1}{\sqrt{(1+zmax)}} )^3 - ( 1 - \frac{1}{\sqrt{(1+z_0)}} )^3)$

where $k$ = 575 Gpc$^3$ per sr.

Summing the contributions of individual sour\-ces,
and normalising by area
(Tabs. 1, 2), we obtain space densities of $0.65 \pm 0.18$,
and $1.75\pm0.36$, quasars per Gpc$^{3}$ at $z \approx 4.2$
and $z \approx 2$ respectively.

Correcting for completeness ($77\pm5$)\% for the $Q_{z4RL}$ sample,
multiplied by $60\%$ for $S \le 1.5$ mJy),
(92$\pm$1.4)\% for the $Q_{z2RL}$ sample) the space densities become:

\indent
$\rho_{z4RL} = 0.99\pm$ 0.28 Gpc$^{-3}$

\indent
$\rho_{z2RL} = 1.90 \pm$ 0.47 Gpc$^{-3}$

\noindent
with ratio:

\indent
$\rho_{z2RL}/\rho_{z4RL} = 1.9 \pm 0.7$.

In calculating the errors, we also took into account possible
systematic shifts of 0\fmag1 in the magnitude scale.

\medskip

Had we adopted the currently more fashionable cosmological {\it concordance
model} with $\Omega_\Lambda=0.8$, $\Omega_{\rm matter}=0.2$ and
H$_0$ = 70 km s$^{-1}$ Mpc$^{-1}$,
the ratio of the comoving volumes $V(z=4)/V(z=2)$ would have been higher
by a factor $\approx 1.25$.
The derived radio and optical
luminosities change by different amounts in the two redshift bins.
E.g. $M_{AB}= -26\fmag9$ in the Einstein -- de Sitter cosmology
becomes in the $concordance ~model$ $M_{AB} = -27\fmag16$ at $z$ = 2, and
$M_{AB}=-27\fmag34$ at $z$ = 4. The same applies to the radio luminosities.
To maintain comparable luminosities in the
two samples, eight of the $z$ = 2
quasars have to be omitted.
The density ratio $\rho_{z2RL}/\rho_{z4RL}$
then decreases by a factor $\simeq 0.81$.

\medskip

\section{The space density of optically-bright quasars at $ z \approx 4$}

In the previous section we measured the space density of optically bright
(M$_{AB}$(1450\AA)$ < -26\fmag9$)
radio-loud ($\log P_{1.4} > 25.7 $) quasars at $z \approx 4$.
At this redshift the total space density of all optically-bright quasars
(i.e. radio-loud + radio-quiet) is still poorly known, although SDSS
should soon yield an accurate value.
Here we obtain it from the space density of the $Q_{z4RL}$ sample
and from an accurate new measurement of the fraction  of optically-bright
quasars with  $\log P_{1.4} > 25.7$ (radio
loud).\footnote {The definition of radio-loudness is rather arbitrary.
We actually define as `radio loud' any quasar detected
in FIRST, i.e. $S_{1.4GHz} \ge$ 1 mJy, which corresponds to
log $P_{1.4GHz} \ge 25.7$ at the median redshift of the $z$ = 4 sample.
Gregg et al. (1996) use the very similar criterion log $P_{1.4GHz} > 25.5 $.}
We indicate this fraction  as $G_{RL}$.

\subsection{The radio/optical fraction of quasars at $ z \approx 4$}

From Djorgovski's (2001) list of $z >$ 4 quasars
we extracted a sample of 191 optically-selected quasars
lying in the area of sky covered by the FIRST survey.
For 22 of these we found  a FIRST radio source within 3$''$.
The number of chance
coincidences will be negligible (the number of candidates does not change
if the search radius is increased to 10$''$),
i.e. all of these are radio-loud quasars.

We are interested in the luminosity range covered by
the $Q_{z4RL}$ sample,
$M_{AB}$(1450  \AA)$ \le -26\fmag9$.
 From
the reported magnitudes, which are APS or APM POSS-I
or POSS-II R,
we computed M$_{AB}$(1450 \AA) for all 191 objects,
with estimated rms and systematic errors both
$\leq 0\fmag3$.
The magnitude uncertainties are
unimportant for our purposes, because we will see that
the fraction of radio loud quasars, $G_{RL}$,
changes little over a broad range of optical luminosity.
Applying the correction factor for the incompleteness of the FIRST
(see Sect. 2.3) we obtain:

$ G_{RL}(M_{AB}({\rm 1450\AA}) < -26\fmag9) = (13.3 \pm 4.8) \%$

$ G_{RL}(M_{AB}({\rm 1450\AA}) > -26\fmag9) = (13.4 \pm 3.7) \%$

i.e. there is no evidence that $G_{RL}$ depends on
optical luminosity.
The mean value is:

$ G_{RL}(-25^m \ga M_{AB}({\rm 1450\AA}) \ga -29^m) = (13.4 \pm 3.0) \%$.

This is consistent with the less accurate estimate
(12 $\pm$ 6) \% by Stern et al. (2000).

In the next paragraph we will use this ratio and the space
density of the $Q_{z4RL}$ loud sample to find the space density of
{\it all quasars} at $z \approx 4$ with $M_{AB} < -26\fmag9$.

\subsection{The luminosity function of optically
bright quasars }

\begin{figure}
\epsscale{.9}
\plotone{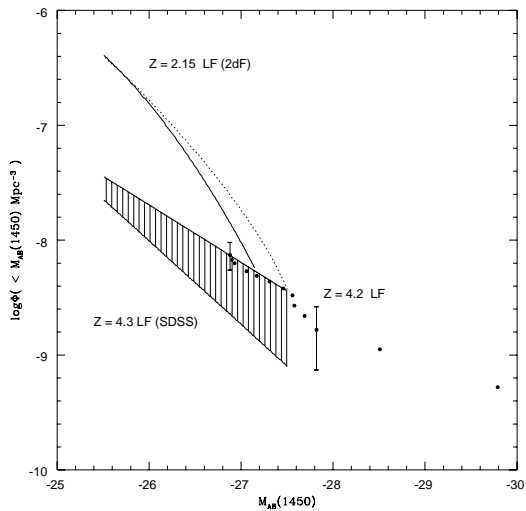}
\epsscale{1.0}
\caption{The cumulative luminosity function
of optically-luminous quasars
at $z \sim $ 4.2, derived in Sect. 4.2
(filled dots, 2 sample error bars shown),
compared with SDSS results for $z =$ 4.3
(shaded area, $\pm 1\sigma$ confidence level).
The solid curve is the cumulative luminosity function obtained by integrating
the 2.0 $< z <$ 2.3 luminosity function from the 2dF survey
(Boyle et al. 2000), the dashed curve is their global model best fit.
 \label{fig4}}
\end{figure}

Above we measured a radio-loud fraction
$G_{RL}  = (13.4 \pm 3.0) \% $ for quasars at $z \sim4$.
We can therefore calculate the
space densities of {\it all quasars} at this $z$,
as a function of $M_{AB}$(1450\AA), simply by
dividing the corresponding space densities of the
radio-loud quasars by $G_{RL}$ e.g.:

$\rho(M_{AB}({\rm 1450\AA}) \le -26\fmag9) = (7.4 \pm 2.6$) Gpc$^{-3}$

and

$\rho(M_{AB}({\rm 1450\AA}) \le -27\fmag5) = (3.8 \pm 1.7$) Gpc$^{-3}$


These numbers are consistent with
the results of Fan et al. (2001b) at $z$ = 4.3 (Fig. 4).
Their maximum-likelihood luminosity function gives a density
of $4.1 ^{+3.8}_{-2.1} Gpc^{-3}$
at $M_{AB}(1450\AA) \le -26.9^m$.

In Fig. 4 we compare our cumulative quasar luminosity function
with that from 2dF
at $z \approx 2$ (Boyle et al. 2000).
We use the actual data (their Fig. 4), rather than their
pure-luminosity-evolution model fit, which
tends to overestimate the space densities at the bright end
(both are shown in Fig. 4).
In order to make this comparison we have converted their $M_B$
to $M_{AB}$(1450 \AA) assuming $M_{B} =
M_{AB}({\rm 1450 \AA}) + 1.2 \alpha_{opt} + 0.12$ (Kennefick et al. 1995) and
adopting $\alpha_{opt} = -0.5$ (Boyle et al. 2000).
In estimating the errors we allow also for possible systematic differences
of 0.1 in the magnitude scales.
The 2dF space density at $z \approx 2$ is:

\indent
$\rho(M_{AB}({\rm 1450 \AA}) \le -26\fmag9) = (13.5 \pm 4$)  Gpc$^{-3}$

\noindent
Thus the ratio between the space  densities, for M$_{AB}$(1450\AA)$<
-26\fmag9$, at $z \approx 2$ and at $z\approx 4$  is:

$\rho_{z2}/\rho_{z4} = 1.8 \pm 0.8$

\noindent
If we extrapolate the z = 2 luminosity function to brighter magnitudes we get:

\indent
$\rho(M_{AB}({\rm 1450 \AA}) \le -27\fmag5) \approx  2.0 ~Gpc^{-3}$

\noindent
and the ratio is less than 1 at $M_{AB}({\rm 1450 \AA}) \le -27\fmag5$.

Had we used the pure-luminosity-evolution model the densities would
 have been larger by a
factor $\approx 1.7$ and $\approx 2.0$ and the density ratios would have
been $\approx 3.0$ and $\approx 1$.

The change in space-density ratio with optical luminosity is evident
in Fig. 4.

\section {Discussion}

We have shown that the space density of
optically-luminous quasars, M$_{AB}$(1450\AA)$ < -26\fmag9$, declines
moderately,
if at all, from  $z \approx$ 2
to $z \approx$ 4, contrasting sharply with the steep decline (factor $\sim$ 10)
found for samples including lower luminosity quasars
(Shaver et al. 1996, 1999,
Warren et al. 1994, Schmidt et al. 1995, Kennefick et al. 1995).

We stress that the quasars in our sample come from the bright tail
of the optical luminosity function.
The difference between our result and those previously reported is
due to the different
cosmological histories of quasars of different optical luminosity,
the space density of the highly-luminous
ones having peaked at earlier epochs than that of the less luminous ones.

This abundance of very bright quasars at $z \approx 4$ has already
been noted
by Fan et al. (2001b), who showed that their luminosity function
at $z\simeq 4$ is flatter than that at lower redshifts (note that this
is not readily-apparent in their Fig. 4,
where the $z \approx 2$ 2dF-luminosity
function is not reported correctly).
Our z = 4 cumulative luminosity function is broadly consistent with that
of Fan et al. (2001b) for SDSS, but extends it by about one decade in optical
luminosity.

The relatively high space density of optically-luminous
quasars at high redshift,
and the consequent existence of
super-massive black holes at very early stages of the evolution of
the universe,
yield non-trivial constraints on models
for structure formation (e.g.: Haehnelt \& Rees, 1993; Granato
et al. 2001; Haiman \& Loeb 2001).

\section{Conclusions}
We sought $z \ge 3.8$ quasars identified with FIRST radio sources
$S_{1.4GHz} > $ 1 mJy
(irrespective of radio spectral index), i.e.
30 times fainter than previous searches,
and built a   sample of 13 optically-bright radio-loud
quasars (M$_{AB}$(1450\AA)$\le -26\fmag9$,
$\log P_{1.4 GHz}({\rm W\,Hz^{-1}}) \ge 25.7$) with  $3.8 < z < 4.5$.
Our conclusions are:

\begin{itemize}
\item{The space density of quasars meeting the above luminosity criteria
is ($1.0 \pm 0.3$) $\rm Gpc^{-3}$}.

\item{From a comparison sample at $z \approx 2$
we find that the space density of radio-loud quasars
declines between
$z \approx 2$ and $z \approx 4$ by a factor 1.9$\pm$0.7.}

\item{We obtained an improved measurement of the radio-loud fraction of
quasars in the luminosity range $\log P_{1.4}({\rm W\,Hz}^{-1}) > 25.7$:
$G_{RL} = (13.4 \pm 3)\%$, nearly independent of optical luminosity, in the
range $-25^m \ga M_{AB}({\rm 1450\AA}) \ga -29^m$.}

\item{Using the above value of $G_{RL}$,
we calculate the  space density of optically-selected quasars with
$M_{AB}$(1450\AA)$ < -26\fmag9$, to be
(7.4 $\pm$ 2.6) Gpc$^{-3}$.
Using the 2dF quasar luminosity function,
we find a ratio $\rho_{z2}/\rho_{z4} = 1.8 \pm 0.8$, consistent with
that for the  radio-loud quasars,
but significantly different from the generally quoted ratio $\sim 10$ found
for samples including lower-luminosity quasars.}

\item{The moderate decline of quasar space density from $z = 2$ to $z = 4$ may
halt altogether at the very bright end of the optical luminosity function,
as, to some extent, is apparent in results from SDSS (Fan et al. 2001b).}

\end{itemize}

\section{Appendix : Absolute magnitude computation}

\begin{figure}[h]
\epsscale{0.9}
\plotone{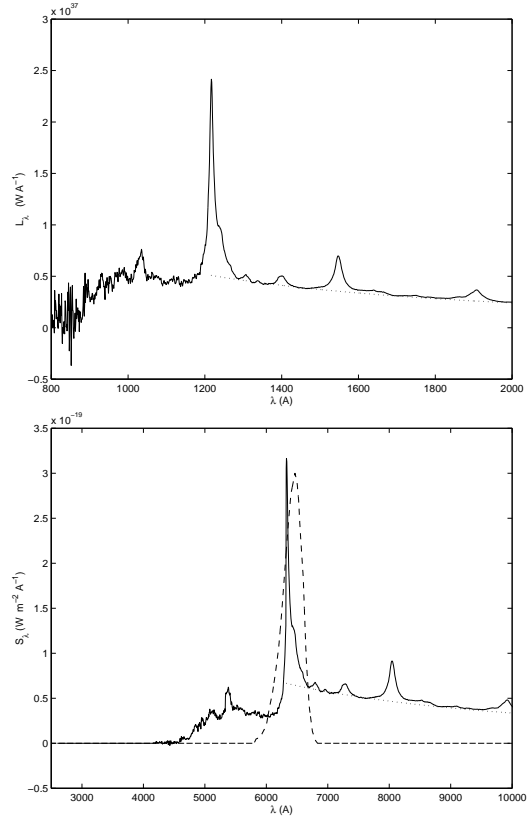}
\epsscale{1.0}
\caption{The top panel shows the FBQS composite quasar spectrum
(Brotherton et al. 2001), overplotted with
a power law with spectral index $\alpha_{\nu} = -0.5$ (dotted curve).
The bottom panel shows the quasar template (solid curve)
in the observer's frame for
$z$ = 4.2, after correction for the Madau (1995) transmission.
The POSS-I $E$ response is shown as a dashed curve.
 \label{fig5}}
\end{figure}
\begin{figure}[h]
\epsscale{0.9}
\plotone{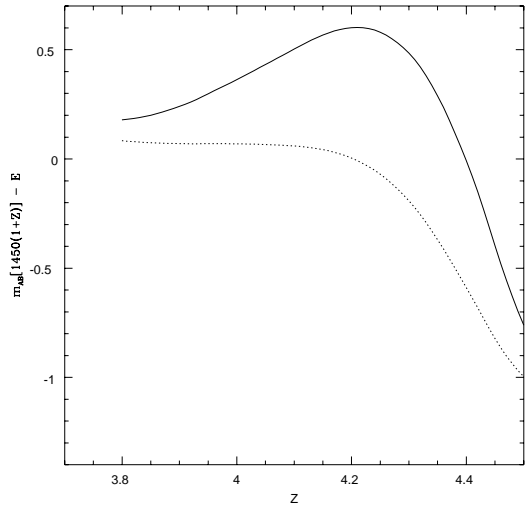}
\epsscale{1.0}
\caption{The $K$-correction between apparent magnitude m$_{AB}$[1450(1+z)] and
apparent $E$ magnitude (POSS-I plate+filter response).
The solid line corresponds to Ly$\alpha$+NV equivalent width 70 \AA,
the dotted line to
15 \AA.
 \label{fig6}}
\end{figure}

We obtained the continuum absolute magnitudes
of the quasars, $M_{\rm AB}(\rm 1450 \AA)$ using:

\vspace{0.2cm}\noindent
 $ M_{\rm AB}({\rm 1450 \AA}) = E - 43.264 - 5 {\rm log}(d_L(z))$

\vspace{0.2cm}\noindent
$- 2.5 {\rm log} \int S_\lambda^{\rm Vega}(\lambda)T(\lambda) d\lambda
- 2.5 {\rm log}(1+z)$

\vspace{0.2cm}\noindent
 $ + 2.5 {\rm log} \int {S_\lambda^Q(\lambda / (1+z)) \over
S_\lambda^Q(1450 \AA)} T(\lambda) d\lambda ~(1)$

\vspace{0.3cm}\noindent
$E$ is the APS POSS-I  magnitude corrected for
Galactic extinction.
AB apparent magnitudes are defined by $m_{\rm AB}
(\lambda) = -2.5 {\rm log} S_\nu(\lambda) -48.60$, where $S_\nu$
is measured in [erg cm$^{-2}$ s$^{-1}$ Hz$^{-1}$] (Oke 1974).
$S_\lambda^{\rm
Vega}$ in [erg cm$^{-2}$ s$^{-1}$ \AA$^{-1}$] was taken from
Fukugita et al. (1996).
$S_\lambda^Q$ is based on the FBQS composite by Brotherton et al.
(2001) and on the
models of Madau (1995)
for intervening absorption.
First, an unabsorbed FBQS composite was created using
the Madau model for $z$ = 2.5.
This was then corrected
for intervening absorption at each redshift using linear
interpolations of Madau's transmission curves for $z$ = 3.5
and $z$ = 4.5 (see Fig. 5 for an example).
$T(\lambda)$ is the response
function of the POSS-I $E$ filter + plate (Evans 1989) and $d_L$ is the
luminosity distance in Mpc.  The $E$ magnitude of Vega,
$E$(Vega) = 0\fmag05, was obtained from the relation $E-R$ vs
$V-R$ by Humphreys et al. (1991), using $V$(Vega) and
$R$(Vega) from Fukugita et al. (1995).

\vspace{0.3cm} \noindent
The combined Ly$\alpha$+NV rest-frame equivalent width (EW) of the FBQS
composite is 70 \AA\ at $z$ = 4.2,
close to the mean, 75 \AA\ (rms $\sim$ 30 \AA), for our $z$ = 4 sample.
The
Ly $\alpha$ line of our quasar
0918+0636 is heavily extinguished, EW(Ly $\alpha$ + NV)$\le$ 15 \AA.
For this quasar the emission-line contour of the template was scaled
to EW = 15\AA.

The relation between $m_{\rm AB}
[1450(1+z)]$ and $E$ is shown as a function of redshift
for 3.8 $< z <$ 4.5 in Fig. 6 for the two values of EW.
$E$ magnitude decreases with increasing strength of the emission line,
particularly at $z \sim$ 4.2,
when Ly$\alpha$+NV fall in the middle of the $E$ band.

\acknowledgments
We thank Prof. G. Zamorani for useful discussion and criticism.
RC and JIGS acknowledge financial support from DGES project
PB98-0409-C02-02 and MCYT project AYA 2002-03326. KHM was supported by a Marie-Curie Fellowhip of the European Commission. JH acknowledges funding via
a PPARC studentship.
This research has made use of the APS Catalog of POSS-I, which is
supported by NASA and the University of Minnesota; and of
the NASA/IPAC Extragalactic Database (NED) which
is operated by the Jet Propulsion Laboratory, California Institute of
Technology under contract with NASA.
We thank an anonymous referee for his comments and suggestions which helped
in improving the presentation of the paper.

\end{document}